\documentclass[preprint2]{aastex}
\usepackage{graphicx}
\usepackage{txfonts}
\usepackage{natbib}
\usepackage[scaled]{helvet}
\usepackage{epsfig}
\usepackage{url}
\bibpunct{(}{)}{;}{a}{}{,}
\begin{document}

\title{\bf Occultation of the T Tauri Star RW Aurigae A by its Tidally Disrupted Disk}
\author{Joseph E. Rodriguez$^{1,2}$, Joshua Pepper$^{3,2}$, Keivan G. Stassun$^{2,1}$, Robert J. Siverd$^2$, Phillip Cargile$^2$, Thomas G. Beatty$^4$, B. Scott Gaudi$^4$}

\affil{$^1$Department of Physics, Fisk University, 1000 17th Avenue North, Nashville, TN 37208, USA}
\affil{$^2$Department of Physics and Astronomy, Vanderbilt University, 6301 Stevenson Center, Nashville, TN 37235, USA}
\affil{$^3$Department of Physics, Lehigh University, 16 Memorial Drive East, Bethlehem, PA 18015, USA}
\affil{$^4$Department of Astronomy, The Ohio State University, Columbus, OH 43210, USA}

\shorttitle{Deep Occultation of RW Aur A}

\begin{abstract}
	RW Aur A is a classical T Tauri star, believed to have undergone a reconfiguration of its circumstellar environment as a consequence of a recent fly-by of its stellar companion, RW Aur B. This interaction stripped away part of the circumstellar disk of RW Aur A, leaving a tidally disrupted ``arm'' and a short truncated circumstellar disk. We present photometric observations of the RW Aur system from the Kilodegree Extremely Little Telescope (KELT) survey showing a long and deep dimming that occurred from September 2010 until March 2011. The dimming has a depth of $\sim$2 magnitudes, a duration of $\sim$180 days and was confirmed by archival observations from American Association of Variable Star Observers (AAVSO). We suggest that this event is the result of a portion of the tidally disrupted disk occulting RW Aur A, specifically a fragment of the tidally disrupted arm. The calculated transverse linear velocity of the occulter is in excellent agreement with the measured relative radial velocity of the tidally disrupted arm. Using simple kinematic and geometric arguments, we show that the occulter cannot be a feature of the RW Aur A circumstellar disk, and we consider and discount other hypotheses. We also place constraints on the thickness and semi-major axis of the portion of the arm that occulted the star. 
	
\end{abstract}

\maketitle
\section{\bf{Introduction}}
Classical T Tauri stars (CTTS) are a type of active pre-main sequence stars that show a large excess of infrared (IR) radiation and were first identified by the broad H$\alpha$ emission line widths in their spectra. The large IR flux in the spectra of CTTS is normally attributed to thermal emission from their dusty circumstellar disks. A large excess of ultraviolet (UV) radiation is often observed and is believed to be a result of material in the disk being accreted onto the surface of the star, producing hot spots \citep{Bertout:1988}. Irregular photometric variability that is characteristic of CTTS has a typical amplitude of less than one magnitude and a time-scale of days to weeks \citep{Herbst:1994}. T Tauri stars are known to have significant magnetic activity which can produce cool star spots and enhanced chromospheric emission. Weak-line T Tauri stars (WTTS), which are typically found in the same star-forming clouds as CTTS, were first distinguished from CTTS by having very narrow H$\alpha$ emission widths \citep{Walter:1986, Herbig:1988}. They were then found to be bright x-ray emitters and show little to no UV or near-IR excess emission. These observational results are generally interpreted as indicating that WTTS are no longer accreting and are either diskless or have disks with very little mass in the form of small, hot dust grains \citep{Haisch:2001}. It is generally believed that CTTS evolve into WTTS when the disk is no longer a significant component of the system, as a result of accretion onto the star, planet formation, and dispersal.  

From surveys, it appears that most nearby T Tauri stars are members of close binary systems \citep{Ghez:1993, Leinert:1993, Richichi:1994, Simon:1995, Ghez:1997A} . These companions can influence the stellar environment and affect the stellar properties determined for the primary star if not taken into account \citep{Ghez:1997}. Depending on the parameters of the companion star's orbit, the companion can significantly disrupt the primary star's circumstellar disk. Simulations show that a fly-by of a stellar companion can tidally disrupt the circumstellar disk around a star, leaving behind a truncated, more compact disk and a long tidal arm feature trailing out from the disrupted material \citep{Clarke:1993}. 

Occultations of stars by their disks, though rare, provide a powerful tool for probing the structure of circumstellar disks. To date, only a few long-duration, deep eclipses of young stars have been discussed in the literature. Most such eclipses appear to be periodic, with the occultation attributed to a disk around a stellar companion or occultation of a star by its own disk. One well known example of this is $\epsilon$ Aur, an F0 giant that experiences long and deep eclipses every 27.1 years. The eclipse has a duration of almost two years and a depth of 0.8 - 1.0 magnitude \citep{Carroll:1991}. The eclipse is attributed to a companion star with its own circumstellar disk \citep{Kloppenborg:2010}. KH 15D is another system, discovered in 1995, which experiences complex eclipsing events with changing properties \citep{Kearns:1998}. That system consists of a non-eclipsing binary star pair that is repeatedly occulted on the binary orbital period by the sharp edge of the circumbinary disk. Long-term secular changes in the occultations are caused by the slow precession of the circumbinary disk across our line of sight \citep{Chiang:2004, Winn:2004}. There are a few other examples of periodically occulting systems (see, e.g. \citet{Bouvier:2007,Plavchan:2008, Grinin:2008}).

Some young stars experience occultations that are not periodic (or at least not known to be). In 2007, a pre-main sequence star (2MASS J14074792-3945427) exhibited a long and extremely deep occultation \citep{Mamajek:2012}. This eclipse was observed by the SuperWASP photometric transit survey \citep{Butters:2010} and the ASAS photometric survey \citep{Pojmanski:2002}. The eclipse lasted $\sim$54 days and had a $\sim$4 mag maximum depth. The cause of the eclipse is thought to be a circumplanetary disk, in the process of formation, analogous to the rings around Saturn. Photometric variability was seen during the eclipse and attributed to gaps in the large ring system around the hypothesized planet \citep{Mamajek:2012}. 

Such systems provide insight into the nature of proto-planetary and circumstellar environments and can be used as tools to probe the structure and composition of circumstellar disks. In this paper we present new observations of the bright system, RW Aur, indicating a long and deep occultation (Figure \ref{fig_full_LC}) somewhat similar to the one seen by \citet{Mamajek:2012}. To our knowledge this paper is the first analysis of the event, which was first reported by Aleks Scholtz\footnote{From Aleks Scholz, A huge eclipse in the young star RW Aur. http://dx.doi.org/10.6084/m9.figshare.92169. Retrieved 21:01, May 13, 2013 (GMT)}. We interpret the event as an occultation of RW Aur A by a portion of its tidally disrupted disk.

The paper is organized as follows. We introduce the known characteristics of the RW Aurigae system in \S2, illustrating the complex stellar environment. In \S 3, we describe the photometric observations, and then discuss the photometric properties of the data in \S 4. In \S5, we present several interpretations of the light curve and discuss their plausibility. We summarize our results and conclusions in \S6.

\section{\bf Characteristics of the RW Aur System}
The RW Aur system ($\alpha$ = 05h 07m 49.566s, $\delta$ = $+30^{\circ}$ 24$\arcmin$ 05.18$\arcsec$) is at a distance of 142 $\pm$ 14 pc \citep{Wichman:1998}, and has a complex stellar environment. It is a binary system, comprising one CTTS (RW Aur B) orbiting another CTTS (RW Aur A) \citep{Duchene:1999}. The primary component (RW Aur A) has a spectral type between K1 and K4 \citep{Petrov:2001}, $V$ $\sim$10.5 \citep{White:2001} and an inferred mass of $\sim$1.3-1.4 $M_{\sun}$, determined by comparing broadband near IR observations to pre-main sequence models \citep{Ghez:1997, Woitas:2001}. The A component shows clear IR excess indicative of a circumstellar disk, with a mass of $M_d$ $\sim$3 $\times$ $10^{-4}$ $M_{\sun}$ \citep{Williams:2006}. This disk radius is calculated to be $\le$ 57 AU, making it one of the smallest detected around a T Tauri star \citep{Cabrit:2006}. RW Aur A exhibits some extreme features, including an accretion rate of 2-10 $\times$ $10^{-7}$  $M_{\sun}$ yr$^{-1}$, one of the highest known for a T Tauri star, \citep{Hartigan:1995} and also bright bipolar jets \citep{Woitas:2002}. \citet{Lopez-Martin:2003} determined that the bipolar jets coming from RW Aur A are inclined to our line of sight by $46^{\circ}$ $\pm$ $3^{\circ}$, using the ratio of the proper motion and the radial velocities. These jets extend out at least 145$\arcsec$ from RW Aur A and contain knots of emission out to $\sim$100$\arcsec$ \citep{Mundt:1998, Woitas:2002}. Transverse velocity shifts have been detected for this jet, which indicate that it may be rotating and that the jet is originating from the disk surface within 1.6 AU from the star \citep{Cabrit:2006,Woitas:2005}

RW Aur B is 1.5\arcsec ($\sim$200 AU) \citep{Duchene:1999, Cabrit:2006} from RW Aur A and has an inferred mass of 0.7-0.9 $M_{\sun}$ \citep{Ghez:1997,Woitas:2001}. It is a late K star (K6 $\pm$ 1 \citep{White:2004}) with $V$ $\sim$13.7 \citep{White:2001}. \citet{Ghez:1993} found faint K-band emission 0.12\arcsec\ from RW Aur B and interpreted this to be a third stellar component to the system, RW Aur C. Other authors have also claimed the presence of a third component in the system, but clear evidence is lacking. Furthermore, later observations failed to confirm the third component \citep{Ghez:1997}.

The RW Aur system has been found to vary in many observational characteristics. \citet{Beck:2001} found that the RW Aur system showed peak-to-peak variations of 2-3 mag on timescales of months using Harvard photographic observations from 1899 to 1952 (See Figure 3 of \citet{Beck:2001}). The standard deviation of these variations is $\sim$0.7 mag. \citet{Petrov:2001} observed periodic variability in the radial velocity of RW Aur A of 2.77 days as well as periodicity in the $U-V$ and $B-V$ colors of 2.64 days. The combined RW Aur system shows an overall non-periodic, photometric variability that \citet{Herbst:1994} determined arises from the A component. The erratic variability has been attributed by different authors to the accretion of the disk onto the star or circumstellar extinction from strong disk winds \citep{Herbst:1994, Petrov:2007}. These are discussed in more detail in \S 4.1.

\citet{Cabrit:2006} conducted millimeter observations of the RW Aur system using the IRAM Plateau de Bure Interferometer (1.3 mm, 2.66 mm). They find in the 1.3 mm observations that RW Aur A has an outer truncated circumstellar disk extending out 41 - 57 AU in radius, inclined by $45^{\circ}$ - $60^{\circ}$ to our line-of-sight, and what appears to be a large, tidally disrupted trailing ``arm'' that is wrapped around the star (see Figure 1.a from \citet{Cabrit:2006}). This arm-like feature has a three-dimensional-length of $\sim$600 AU, and is almost certainly the result of a recent stellar fly-by of RW Aur B \citep{Cabrit:2006}. This fly-by also caused the RW Aur A disk to be truncated near the periastron separation. While a large portion of the arm is redshifted, indicating it is wrapped around and behind the A component, there is a small portion that is connected to the northeast side of the RW Aur A disk which is blue-shifted by up to 3.1 km s$^{-1}$ relative to the star. This feature resembles the destructive outcome of a coplanar eccentric fly-by similar to simulations from \citet{Clarke:1993}, where all material outside the periastron distance is fully disrupted and drawn out into a coherent tail in the direction of the companion's orbit. The RW Aur system therefore provides an excellent example of a T Tauri disk that has experienced a recent dynamical disruption.

\section{\bf Photometric Observations}
Several photometric surveys have observed RW Aur over both short and long time scales going back to 1899. The light curve data are shown in Figures \ref{fig_full_LC} and \ref{fig_full_LC_Zoom}. 

\subsection{Archival Data}

The Wide Angle Search for Planets (SuperWASP) is a wide field photometric survey designed to detect transiting extrasolar planets over a large fraction of the sky. SuperWASP observed RW Aur for one shortened season in 2004 and then two later seasons in 2006 and 2007. The SuperWASP public archive data is described in detail in \citet{Butters:2010}. The SuperWASP observations have a cadence of a few minutes in the $V$ band and do not resolve the RW Aur system, thus the light curve incorporates light from all system components. The median error for SuperWASP is $\sim$0.01 Mag.

The American Association of Variable Star Observers (AAVSO) is a non-profit organization dedicated to the goal of understanding variable stars. The AAVSO archive contains observations of RW Aur going back to 1937, with the observations increasing in cadence around 1954. The AAVSO data consist of $V$ band and visual observations. Only some of the AAVSO data have corresponding uncertainties reported. 

Wesleyan University's Van Vleck Observatory has monitored many known T Tauri stars with the 0.6 meter Perkin Telescope. The resulting observations are included in a public archive of ${UBVRI}$ photometry. RW Aur A was observed from January 1965 until October of 1994, with varying frequency. A detailed description of the archive is described in detail in \citet{Herbst:1994}.

The AC and AM photographic plate series at the Harvard College Observatory have observations of RW Aur from 1899 to 1952, resulting in 162 observations. The photographic observations were obtained using the 1.5 inch Cooke lens, corresponding to a plate scale of 600\arcsec/mm. The B band magnitudes were estimated by analyzing over 150 archival plates and comparing them to the known B magnitudes of nearby stars (Not shown in Figure 1, see Figure 3 of \citet{Beck:2001}).

\subsection{KELT-North}

The Kilodegree Extremely Little Telescope (KELT-North) is an ongoing survey, searching for transiting planets around bright stars ($V$ = 8-10). KELT-North uses a Mamiya 645 80mm f/1.9 lens, and has a 42 mm aperture and a large field of view ($26^{\circ}$ $\times$ $26^{\circ}$) with a plate scale of 23\arcsec per pixel \citep{Pepper:2007}.

RW Aur is located in KELT-North Field 04, which is centered on ($\alpha$ =  5hr 54m 14.466s, $\delta$ = $+31^{\circ}$ 44$\arcmin$ 37$\arcsec$). KELT-North observed this field from October 10, 2006 to September 23, 2012,  obtaining 8,001 images. The data were reduced using a heavily modified version of the ISIS software package, described further in \S2 of \citet{Siverd:2012}\footnote{Much of the software is publicly available at the following address:  http://verdis.phy.vanderbilt.edu}. The observations are in a broad $R$-band filter, with a $\sim$15 minute cadence and the typical error is less than 0.04 Mag. The KELT-North observations also fail to resolve the RW Aur system. The KELT-North data is presented in Table 1.

\begin{table}[ht]
\caption{KELT-North photometric observations of RW Auriga}
\begin{tabular}{ | c | c | c |}
\hline
BJD$_{\! {\rm  TDB}}$   &         KELT $V$ Mag\tablenotemark{a}    &    Poisson Errors\tablenotemark{b} \\
\hline

2454035.812879   &    10.326  &    0.008 \\
2454035.817502   &    10.327  &    0.008 \\
2454035.822123   &    10.324  &    0.008 \\
2454035.826745   &    10.319  &    0.008 \\
2454035.831367   &    10.311  &    0.008 \\
2454035.835989   &    10.294  &    0.007 \\
2454035.840612   &    10.286  &    0.007 \\
2454035.845235   &    10.281  &    0.007 \\
2454035.849855   &    10.286  &    0.007 \\
2454035.854477   &    10.283  &    0.007 \\
\hline
\end{tabular}
\footnotesize{{\bf Notes.} Table 1 is published in its entirety in the electronic edition of this paper. A portion is shown here for guidance regarding its form and content.

$^{\rm a}$Relative KELT Instrumental magnitude corrected to Johnson V-magnitude.  Absolute accuracy to $\sim$0.2 mag.  Relative accuracy to $\sim$0.04.

$^{\rm b}$Poisson errors to instrumental KELT magnitudes.  True per-point magnitude errors must fold in 0.036 mag systematic errors.}
\end{table}

\section{Results: Variability Before and During the Dimming Event}
Here we discuss the variability characteristics of the deep dimming event. We also describe the general photometric variability of RW Aur A. In Section 5 we discuss the dimming in the context of an interpretation in which the star has been occulted by a portion of its tidally disrupted disk.We focus our analysis on four data sets: KELT-North, AAVSO, SuperWASP, and Wesleyan (Van Vleck).

\subsection{Pre-Dimming Variability}
\begin{figure*}[!ht]
\centering\epsfig{file=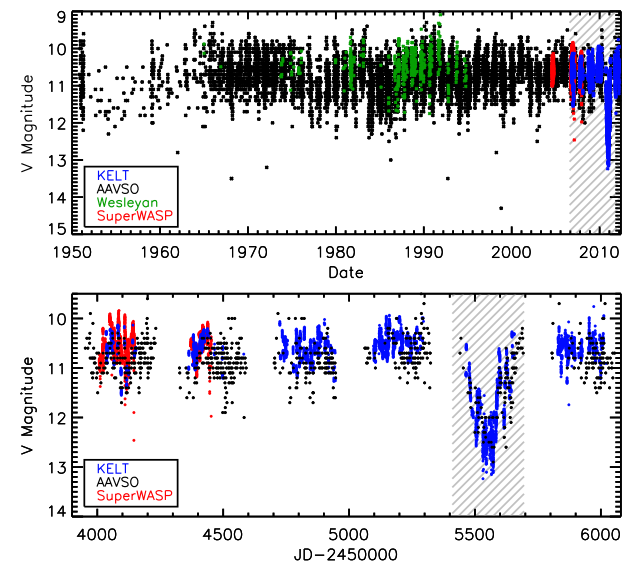,clip=,width=0.95\linewidth}
\caption{(Top) AAVSO (Black), KELT-North (Blue), SuperWASP (Red) and the Wesleyan Van Vleck (Green) light curves of RW Aur from 1950 to 2012. The KELT and SuperWASP light curves do not resolve the A and B components. The shaded region in the upper plot corresponds to the six KELT seasons which is shown in the bottom plot. (Bottom) The KELT-North, SuperWASP and AAVSO light curves plotted for the six KELT-North observing seasons. The dimming of the star is seen from late 2010 through early 2011 in the KELT-North and AAVSO light curves and is centered on early January of 2011. The median KELT and SuperWASP errors are $\sim$0.01 Mag. For better visualization of the true nature of the data, the errors are not plotted. The shaded region in grey is the location of the main dimming event is seen in late 2010 through early 2011. A zoom in of this region is shown in Figure \ref{fig_full_LC_Zoom}. }
 \label{fig_full_LC}
\end{figure*}

CTTS can have erratic photometric variations on the time scale of a few days \citep{Herbst:1994}. The light curve for RW Aur shows this photometric variability and it is observed in all four data sets (Figure \ref{fig_full_LC}). The variability seen in the four data sets for RW Aur has a peak-to-peak amplitude of 1-2 mag, on the timescale of days to weeks, with a standard deviation of $\sim$0.4 mag. 

Two possible explanations for the non-coherent photometric variability are circumstellar extinction and varying accretion onto the surface of the star. \citet{Herbst:1994} argued convincingly that the non-coherent variability is caused by irregular accretion onto the surface of RW Aur A, creating hot spots that rotate into and out of view. \citet{Herbst:1994} performed an analysis of the ${UBVRI}$ photometric observations finding that the photometric variations are roughly divided evenly (with relatively large amplitude) between positive and negative excursions relative to a well defined median level. This is a defining characteristic of accretion-driven variability, as opposed to circumstellar obscuration-driven variability which tends to produce mainly dimming of the star, or flare driven-variability which tends to produce mainly brightening events. Importantly, the H$\alpha$ flux appears to be correlated with the brightness of the star at all wavelengths \citep{Herbst:1982}. This is very strong evidence in favor of an interpretation in which the photometric variability of RW Aur A is principally connected to variations in its accretion rate. \citet{Petrov:2007} and \citet{Grinin:2004} suggest that the photometric variability is the result of a strong disk wind lifting material from the disk across the star causing circumstellar extinction. Such disk winds are a known feature of CTTS but at odds with the accretion variability interpretation of \citet{Herbst:1994}. 

Some accreting stars can show periodic modulations in their light curves due to hot accretion spots on the stellar surface rotating in and out of view on the stellar period. To look for such periodicity we use a Lomb-Scargle periodogram as presented in \citet{Lomb:1976, Scargle:1982,Press:1989}. This Lomb-Scargle method performs spectral analysis on unevenly sampled time series data and allows us to effectively identify weak periodic signals. We performed this analysis on the KELT, AAVSO, SuperWASP and Wesleyen photometric data. 

\begin{figure*}[t]
\centering
\centering\epsfig{file=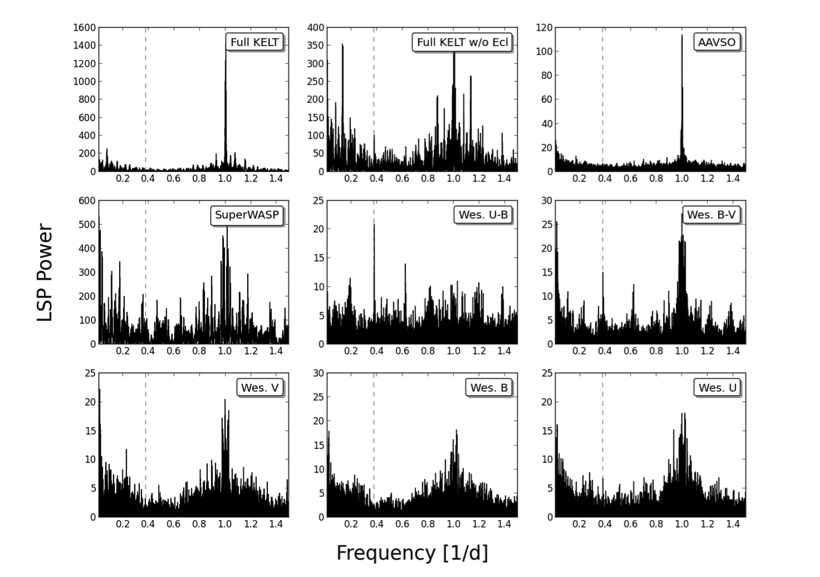,clip=,width=1.0\linewidth}
\caption{The resulting periodograms from the Lomb-Scargle analysis of the AAVSO, KELT-North, SuperWasp and Van Vleck photometric data. The vertical dashed line represents a 2.7 day periodicity that has been previously reported as half the rotation period of the star. We recover this period in the Wesleyan $U-B$ light curve data. The features seen at 1.0 c/d are the (high-frequency) aliases of long-term variation caused by diurnal sampling.
 \label{fig_LS}}
\end{figure*}

We do not see any significant periodicity in the full KELT-North RW Aur light curve, nor in the KELT-North light curve with the dimming in late 2010 removed aside from the 1 day diurnal sampling effect. The AAVSO and SuperWASP light curves also do not show any periodic signal. We do, however, confirm a $\sim$2.7 day periodic variability seen by \citet{Petrov:2007} in the Weslyan University ($B-V$) and ($U-B$) curves. The KELT and SuperWASP light curves each exhibits a small peak between 2.69
and 2.72 days. Though not individually statistically significant, their presence does lend additional credence to this signal. A periodicity at twice this value ($\sim$5.5 days) has been reported as the rotation period of the star \citep{Petrov:2001, Gahm:1999}. We are unable to identify this periodicity in the individual $U$, $B$, or $V$ Weslyan University light curves (Figure \ref{fig_LS}), which could be the result of the rotational modulation signal being masked by the stochastic accretion variability.

While the periodicity analysis does not shed much new light on the physical origin of the variability, the overall available evidence appears to favor an interpretation in which the photometric variability is tied to variations in the accretion rate. Spectroscopic monitoring observations obtained during the deep dimming event corroborate this interpretation also, as discussed in the next section.

\subsection{2010-2011 Dimming}
\begin{figure*}[!ht]
\centering\epsfig{file=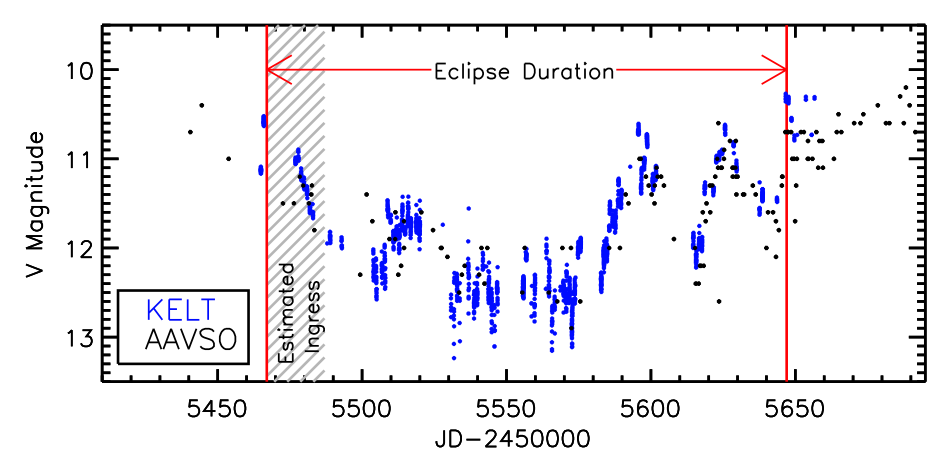,clip=,width=1.0\linewidth}
	\caption{KELT-North (black) and AAVSO (blue) light curves zoomed in on the eclipse. In grey dashed highlights is the estimated ingress of 20 days and the two red vertical lines mark the estimated eclipse duration. The KELT observation have an error of 0.04 Mag while the most AAVSO data do not have reported errors. The faintest points observed during the dimming are near the observational limit of KELT. 
	\label{fig_full_LC_Zoom}}	
\end{figure*}

In late September 2010 the light curve of the RW Aur system became fainter, dropping from a median brightness of $V = 10.4$ to $V = 12.5$ mag (Figure \ref{fig_full_LC}). This decrease lasted for $\sim$180 days, ending in late March 2011. A good deal of structure and variability, on similar time scales as the ingress and egress, 10-30 days, is apparent during the dimming (Figure \ref{fig_full_LC_Zoom}). The peak-to-peak depth of the fading is $\sim$4 mag and has a sustained duration of many months. Therefore, the dimming event has a much deeper and longer duration than the out of event stochastic variations that present a peak-to-peak amplitude of 1-2 mag and a standard deviation of $\sim$0.4 mag on shorter timescales. 

During the dimming, the system's brightness decreases to $V$$\sim$12.5 (Figure \ref{fig_full_LC_Zoom}). For reference, the B component by itself was previously measured to be $V$$\sim$13.7 \citep{Ghez:1997}. Assuming the dimming is only due to a decrease in the brightness of the A component, we calculate that the $V$ magnitude of the A component, during the dimming event, is $\sim$12.9. That corresponds to a decrease in flux of the A component of 91\%, making it $\sim$1.5 times brighter than the B component during the 2010-2011 event. The large amplitude non-periodic short-timescale variability that is so ubiquitous outside the 2010-2011 dimming, appears to significantly diminish but is still present. This is expected if the short-term variability, originating from the A component, is subjected to significant dilution from the combined system with the B component during the dimming. This supports the interpretation that the A component is the source of the non-periodic variability. At the same time, we do observe structure during the dimming, in particular a 1-1.5 mag brightening and then re-dimming toward the end which occurs with a characteristic timescale of $\sim$10 days, very similar to the main ingress timescale. Below we discuss the possibility that this structure during the dimming may imply substructure in the occulting body.

RW Aur A was serendipitously monitored spectroscopically during October through December of 2010 by \citet{Chou:2013}. Although the authors were unaware of the 2010-2011 dimming, their spectroscopic observations by chance coincide precisely with the first half of the deep dimming event, including most of the ingress and the point of maximum depth. Their analysis of 14 different emission lines tracing accretion, infall and outflow, are broadly consistent with both the nature of the line profiles and with the short timescale variability observed by others many years before (see references in \citet{Chou:2013}). In addition, their observed correlations between the various emission lines are consistent with a largely steady magnetospheric accretion model over the course of the observations. They do infer variations in the magnetospheric accretion of order 20 percent, comparable to previous accretion variability measurements for RW Aur A and with the accretion variability interpretation for the general photometric variability of the system \citep{Herbst:1994}. More fundamentally, these serendipitous spectroscopic observations suggest that the accretion behavior of RW Aur A was not connected with the source of the pronounced photometric dimming event.

It is clear that a dimming this large is not present prior to 2010 in the combined light curves from KELT-North, AAVSO, Van Vleck and SuperWASP. Although there is substantial variation seen in the full light curve across 60+ years (Figure \ref{fig_full_LC}), we observe no comparable events in duration or depth. We verify that by examining time spans in the full data set that might appear to represent similar eclipses, but a detailed look shows that those events clearly do not resemble the deep long and coherent event in 2010 to 2011. Since we can place a lower limit on the duration of the fading to be $\sim$180 days from the KELT and AAVSO light curves, we searched for a gap in the combined light curve, large enough, that a similar previous event may have occurred but have been missed. We examine the RW Aur light curve, using all four data sets, and find no gap in the observations greater than 180 days since 1961. Prior to 1961, the only photometric observations of this target are the Harvard photographic observations and a much sparser set of AAVSO observations. Those data span the years 1899 to 1961, with frequent gaps of 200+ days. We can therefore conclude, that if the 2010 to 2011 event repeats it would have a minimum period of $\sim$50 years.

\section{\bf Interpretation and Discussion}
\subsection{\bf Favored Interpretation: Occultation by the RW Aur A Tidally Disrupted Disk}

In this section we present what we regard as the most plausible interpretation of the 2010-2011 dimming event, namely an occultation of RW Aur A by a fragment of the tidal arm that resulted from its disrupted circumstellar disk. This known feature of the circumstellar environment provides the necessary ingredients of a large opaque body, with a sharp edge moving perpendicular to the line of sight, to be able to naturally explain many of the features of the observations. We explore alternate interpretations in \S 5.2, but here we investigate the necessary properties of the occulting body and discuss how well these properties agree with those observed for RW Aur A's tidally stripped arm.

We calculate the key characteristics of the occulter by modeling the dimming as an occultation of RW Aur A by a large body which possesses a sharp edge that is oriented perpendicular to the direction of motion. From the photometric observations, we are able to determine three important features about the occulter: transverse velocity, semi-major axis and width. From the light curves, we estimate the ingress to be between 10-30 days and the total duration to be about 180 days. As a result of the cadence and gaps in observing between the KELT-North data and AAVSO, the values determined for these two parameters are based on a visual inspection of the light curves. We choose to define the end of the dimming event to be after the two large variations at $\sim$5600 and $\sim$5640 (JD- 2450000) (Figure \ref{fig_full_LC_Zoom}) since this type of variability is not seen anywhere in the light curve outside the dimming event and thus can be attributed  to the mechanism responsible for the event. We also observe that the variations during the dimming (including the substructure near the end of the event), all share the same characteristic timescale of the ingress/egress (10-30 days). 

Using the known mass and age of RW Aur A (1.3-1.4 $M_{\sun}$, 4.5-6.2 Myr \citep{Ghez:1997, Woitas:2001}), we refer to the Dartmouth Stellar Evolution Program's (DSEP) young stellar models to find that the radius should be between 1.5-1.7 $R_{\sun}$ \citep{Dotter:2008}. These specific models give a log($T_{\rm eff}$) of 3.67-3.70 that is consistent with the effective temperature determined for RW Aur A by \citet{Liu:2012}. For our calculations, we adopt a mass of 1.35 $\pm$ 0.135 $M_{\sun}$ (error estimate for isochrone fitting) and a stellar radius of 1.6 $\pm$ 0.32 $R_{\sun}$ (conservative error estimate for the stellar evolution models). 

We use the stellar radius and the minimum timescale of the ingress and characteristic variation seen during the occultation (10 days) to calculate the maximum transverse velocity of the occulter to be 2(1.6 $R_{\sun}$)/10 days = 2.58 $\pm$ 0.52 km s$^{-1}$. Furthermore the calculated velocity of the occulter and the total observed duration of the occultation yields a physical width of the occulting body of 2.58 km s$^{-1}$ $\times$ 180 days = 0.27 $\pm$ 0.05 AU. Assuming Keplerian motion and a circular orbit, a velocity of 2.58 km s$^{-1}$ implies a semi-major axis of 180.5 $\pm$ 28.9 AU: 

\begin{equation}
\frac{1.35M_\sun G(10\, {\rm days})^2}{4(1.6R_\sun)^2} \sim 180.5  {\rm AU}. 
\end{equation}

\noindent Using instead the estimated maximum ingress timescale (30 days) we calculate a minimum transverse velocity of 0.86 $\pm$ 0.17 km s$^{-1}$, semi-major axis of 1624.7 $\pm$ 260.1 AU and a occulter width of 0.089 $\pm$ 0.02 AU. At 180 AU, the occulter would be more than three times as far from RW Aur A as the maximum estimated extent of the circumstellar disk (57 AU, see \S2).  Unless the occulter is moving at an extremely oblique angle, these values should be roughly accurate to within a factor of a few. 

\citet{Cabrit:2006} conducted millimeter observations of the RW Aur system and found that the system appears to have undergone a complex interaction with its companion, RW Aur B, that resulted in a short truncated circumstellar disk around RW Aur A (inclined by $45^{\circ}$ - $60^{\circ}$) and a large (600 AU along the spine) tidally disrupted arm that wraps around behind the A component (see \S2). The arm is believed to be mostly wrapped behind the A component, although a small portion appears to be in front. The portion of the arm in front of the A component has a blue shifted velocity relative to RW Aur A of $\le$ 3.1 km s$^{-1}$ which is similar to our calculation of the occulter's transverse velocity. The 0.86 km s$^{-1}$ velocity, for the 30 day ingress calculation, is also consistent since \citet{Cabrit:2006} measured a range of blue shifted velocities in their millimeter observations.

The observations of the tidal feature are consistent with the simulations of a coplanar stellar interaction conducted by \citet{Clarke:1993}. There is no indication that the orbit and disk plane are not in the same plane, which is the most probable configuration. \citet{Bisikalo:2012} found that the orbit of the binary is retrograde to the orbit of the circumstellar disk around RW Aur A. The simulations by \citet{Clarke:1993} for a co-planar, retrograde interaction, not only produce a tidal arm, but also show that the interaction disrupts a significant amount of material out of the disk plane (see Figure 7 from \citet{Clarke:1993}). Therefore it is plausible that material was disrupted out of the disk plane into our line-of-sight even though the system is significantly inclined. 

Although the distance between RW Aur A and the arm is unknown, simulations by \citet{Clarke:1993} show that in an eccentric stellar interaction, the disrupted arm spirals outward from the primary component and its closest point would be outside the extent of the disk. Thus our calculated semi-major axis of $\sim$180 AU (or larger for 30 day ingress) is consistent with simulations of the hypothesized interaction. Without more information about the system configuration during the flyby, we have no definitive knowledge of the full three-dimensional direction of movement for the blueshifted component of the arm. We therefore expect that it is unlikely that the velocity vector is extremely oblique. Thus, we expect that the transverse velocity we calculate is within a factor of a few of the true velocity. Since our calculated velocity is quite similar to the radial velocity seen in the \citet{Cabrit:2006} observations, that congruence supports the conclusion that the disrupted arm (likely a fragment of the full arm) and the occulter are the same body.

The occultation displays a large maximum depth ($\sim$2 mag) and a long duration (180-210 days), with some substructure occurring on a timescale similar to the ingress, 10-30 days (Fig \ref{fig_full_LC_Zoom}). There are large amplitude variations observed near the end of the dimming (5590 - 5650 JD UTC-2450000). These features appear to repeatedly brighten from the maximum dimming depth to the median out of occultation magnitude suggesting that the occulter has substructure and potentially gaps. The faintest points during the occultation are near the edge of KELT's detectability, compromising our ability to characterize any short period variability during the dimming. Within this interpretation, the brightness variations seen during the occultation are due to the substructure of the tidal arm fragment. 

Having made several simple assumptions (sharp leading edge, Keplerian motion, etc.), our calculated properties of the occulter are consistent with the observed properties of the tidal arm from \citet{Cabrit:2006}. The distance between RW Aur A and the occulting body derived in this section ($\sim$180 AU) is comparable to the separation between the RW Aur A and RW Aur B components ($\sim$200 AU). This location implies that the occulting body cannot be a circumbinary object and is probably a fragment of one of the components of the system. Therefore we believe that the best explanation for the occultation mechanism, producing the 2010-2011 occultation, is that a fragment of the tidally disrupted arm crossed in front of RW Aur A. 

\subsection{\bf Alternate Explanations}

We have presented evidence for an interpretation in which the deep, long duration dimming of RW Aur A is due to occultation by its tidally disrupted circumstellar disk. We now explore alternate explanations for these observations.

\subsubsection{Occultation by Stellar Companion}

We can rule out the possibility that the occulter is comparable in size to RW Aur A because the combined ingress/egress timescale would need to be similar to the entire duration of the dimming. If the cause of this event was the result of an eclipse of the A component by a large unseen stellar companion, this would require the companion to have extreme stellar parameters. We can model the system as a large stellar disc passing in front of a smaller one (RW Aur A). This allows us to use the same calculations as in \S5 for the velocity, semi-major axis and diameter (projected width) of the occulting star. The same calculations apply because the eclipsing star's leading edge would be perpendicular to is tangential motion. The occulting star would thus need to be moving $\sim$2.5 km s$^{-1}$ and have a diameter of $\sim$0.27 AU, corresponding to a radius $\sim$58 $R_{\sun}$, a giant star. The star would also need to be dark and cause the large variations at the end of the dimming (Figure \ref{fig_full_LC_Zoom}). From these observed and calculated characteristics, we are confident the occulter is not an unseen stellar companion. 

\subsubsection{Alternate Stellar Parameters of RW Aur A}

We address the assumptions in our calculations of the occulter distance from from RW Aur A. In \S5 we determined that the occulting body must be more than three times as far from the star as the outer edge of the known circumstellar disk. Since that calculation is based on our determination of the linear velocity of the occulter, which in turn depends on the radius of the star, we ask whether our stellar radius estimate (1.6 $R_{\sun}$) could be incorrect. Given the observed maximum extent of the disk of 57 AU \citep{Cabrit:2006} and assuming Keplerian motion, we calculate that RW Aur A would need a radius of at least $\sim$2.75 $R_{\sun}$ for the occulter to be located at the edge of the disk. 

However, a star with a radius 2.75 $R_{\sun}$ would be much more intrinsically luminous. A star with a radius of 2.75 $R_{\sun}$, log($T_{\rm eff}$) of 3.684, and an apparent magnitude of $V$ = 10.4, would be at a distance of $\sim$218 parsecs, which is much larger than the measured distance to RW Aur of $\sim$140 pc \citep{Wichman:1998}. Furthermore, according to the \citet{Dotter:2008} stellar models, a radius of 2.75 $R_{\sun}$ implies a stellar age of $\sim$9.5 $\times$ $10^5$ years. RW Aur is part of the Taurus-Auriga association, where star formation is thought to have first occurred on the outer edge and progressed inward. The youngest estimated age for the Taurus-Auriga stellar association is $\sim$1 Myr for the center of the association; however, RW Aur is located on the outer edge of the region, where star formation is believed to have occurred much earlier and corresponds to an age closer to 10 Myr \citep{Palla:2002}. Since a radius of 2.75 $R_{\sun}$ is not consistent with either the apparent magnitude or age of RW Aur A, it is not likely that the estimated radius is incorrect. 

\begin{figure}[ht]
	\centering\epsfig{file=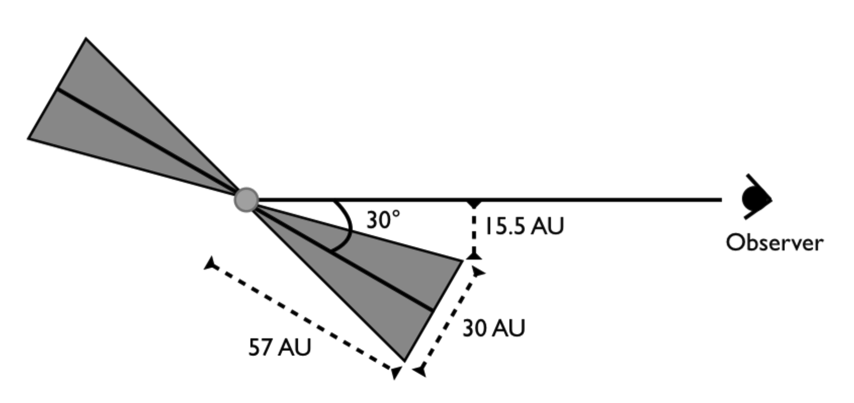,clip=,width=1.0\linewidth}
	\caption{Schematic of the RW Aur A disk geometry, showing to scale, the height required for a feature at the edge of the disk to cause the observed occultation.\label{fig_Disk_Geometry}}	
\end{figure}

\subsubsection{Occultation from Outer Edge of RW Aur A's Circumstellar Disk}

We explore the possibility that a large feature at the edge of the circumstellar disk around RW Aur A has occulted the star. Even though the disk is inclined to our line of sight, pre-transitional disks are not uniformly flat, and tend to flare as a function of semi-major axis. An example of this is clearly seen in \citet{Espaillat:2010} which shows that the height of the disk is 13.8 AU at a semi major axis of 71 AU but only 0.009 AU at 0.1 AU from the star. 

To determine the plausibility of a a feature at the edge of the RW Aur A circumstellar disk causing the occultation seen in late 2010, we model the feature to be in a Keplerian orbit, at the farthest estimated extent of the disk (57 AU) and calculate the additional height required to cross the face of the star. For our model, we consider a conservative disk inclination from \citet{Cabrit:2006} for RW Aur A of  $60^{\circ}$ and a flared disk height of 15 AU above the mid-plane at the outer edge. We assume that the disk flares out linearly to 15 AU. The flared disk height we use is an extreme example of what is seen in other disks \citep{Espaillat:2010}. Using geometric arguments, we calculate that the necessary increase in height for a feature at the disk edge to occult the star is 15.5 AU in addition to the extreme flaring already assumed for the disk (Figure \ref{fig_Disk_Geometry}). 

Using instead the argument of Keplerian motion, a feature at the edge of the disk would have an orbital velocity of $\sim$4.66 km s$^{-1}$, corresponding to a width of $\sim$0.49 AU. This results in the postulated feature with a projected height and width of $\sim$15 AU and $\sim$0.5 AU respectively. A feature with these dimensions would be so tall and thin as to be highly implausible. 
 
\subsubsection{A Warp in the Inner Circumstellar Disk}
RW Aur A appears to have undergone a close interaction with its companion, RW Aur B, in the recent past. Here we consider the plausibility that a warp in the inner part of the RW Aur A disk, caused by the flyby, could cause the dimming seen in 2010-2011. The models by \citet{Clarke:1993} for a retrograde, co-planar orbit show that even though all disk material outside periastron is fully disrupted, the interaction has little effect on the disk interior to this. Therefore, the \citet{Clarke:1993} simulations predict that an interaction of this type would not warp or twist the innermost part of the disk. Other mechanisms such as a misaligned magnetic field, stellar radiation, or planetary companions could cause a warp. The longevity of a warped circumstellar disk is dependent on the mechanism that causes it. 

A misaligned magnetic field, where disk material is funneled onto the stellar surface, would show variations on the timescale of days to a few weeks (e.g., AA Tau; \citet{Bouvier:2003}) because this is the Keplerian timescale in the disk at a distance of a few stellar radii, which is the extent of the star-disk magnetic interaction. The magnetic misalignment effect, such as is seen around AA Tau, could not result in the 180 day dimming of RW Aur A. 

If the warp is caused by the presence of planetary companions, the stability of the warp should be related to the stability of the planets' orbit and therefore the warp should last many orbital periods \citep{Burrows:1995, Mouillet:1997}. \citet{Armitage:1997} showed that radiation-induced warping is possible in high luminosity stars ($L_{*}$ $\gtrsim$ 10$L_{\sun}$) as long as the central star's intrinsic luminosity is much higher than what is created by accretion from the disk. This was an alternate explanation for the warp seen in the disk around $\beta$ Pictoris \citep{Armitage:1997}. 

\begin{figure}[ht]
\centering\epsfig{file=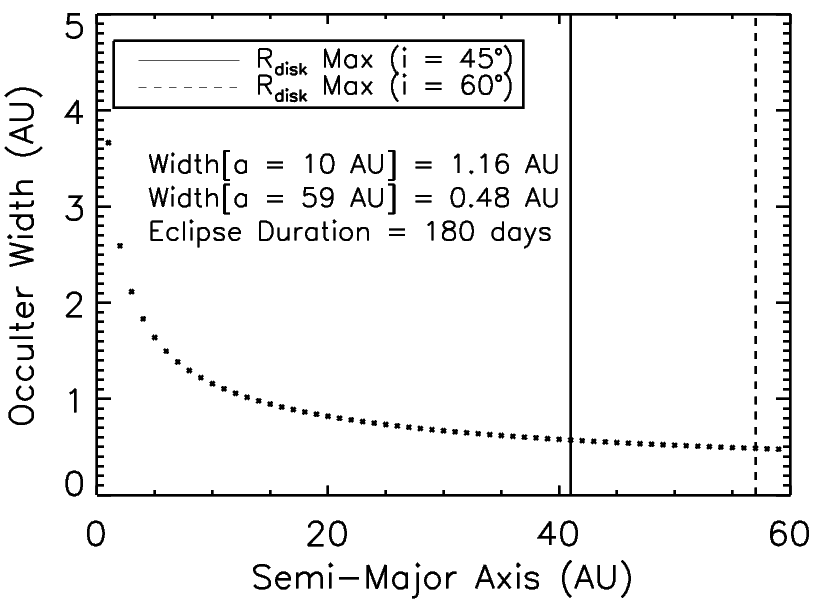,clip=,width=1.0\linewidth}
\caption{Calculated width of the occulting body as a function of semi-major axis for the given observed occultation duration of $\sim$180 days, assuming Keplerian motion. The vertical lines correspond to the maximum disk radius of 41 to 57 AU, based on the two possible inclinations that \citet{Cabrit:2006} estimated for the disk ($45^{\circ}$ and $60^{\circ}$).
 \label{fig_Occultation_Width}}
\end{figure}

To explore the possibility that the occultation is caused by a warp, we adjust our interpretation from \S5.1 and now model the occulter as an opaque body with a sharp leading edge moving across the face of the star at an oblique angle. The obliquity of the leading edge allows the occulter to move at a higher velocity, thus potentially placing it closer to the star. To determine whether a warp is plausible or not, we must determine two key characteristics, the additional height required to cross into our line-of-sight and width of the warp. As we place a warp closer to the star, the orbital velocity increases, which as a result would increase the calculated occulter width (Figure \ref{fig_Occultation_Width}). For a warp to be plausible, it would likely need a larger width then height. Assuming the same scenario as in the previous sub-section, where the disk flares out linearly to a height of 15 AU, we determine through geometric arguments that the only location where the calculated width of the occulter is larger than the line-of-sight height is at a distance from the star of $\le$11 AU. 

This would require the leading edge of the occulter to cross at a highly oblique angle, $\le$15$^{\circ}$ angle relative to the direction of motion. This semi-major axis for the feature corresponds to an orbital period of  $\le$31 years, well within our window of observations and so we should have observed it to repeat. It does not seem plausible that the presence of planets could create and dissipate a warp in less then one orbital period, especially since a planet-induced warp should be stable for longer than its orbital period. 

The accretion rate is too high and luminosity of RW Aur is too low for it to cause a radiation induced warp in its inner disk. Also, the stability of the warp by induced radiation would be very long and should have been observed more than once in 50+ years of constant observation. In both potential disk warping mechanisms (planetary companions and stellar radiation), the lifetime of the warp would be much longer then the orbital timescale and would have been observed more than once. Therefore, we do not believe that a warp is a likely interpretation of the 2010-2011 dimming. 

Finally, we examine the possibility that the circumstellar disk around RW Aur A has eclipsed the star once due to a large precession of the disk. From our calculations, we are confident that the occulter is not located inside the disk around RW Aur A (minimum semi-major axis $\sim$180 AU compared to the maximum radius of the disk, 57 AU \citep{Cabrit:2006}).  We have determined that due to the inclination of the disk, any warp would need to be extremely large to cross our line of sight. This would also require an extreme precession of the circumstellar disk to force it into our line of sight. Thus precession is not a plausible explanation.

Given the extreme disk distortions required by these scenarios, we can conclude that the circumstellar disk around RW Aur A is not a likely explanation for the 2010-2011 dimming.

\subsubsection{UXor Variation}
UX Orionis stars are a class of pre-main sequence stars, typically Herbig Ae/Be objects, that experience large dimming events, sometimes described as Agol-type minima, in the $V$ band of up to 3 mag, lasting on timescales of days to months. These events manifest as sudden drops in the visual brightness followed by a slow recovery. The minima events are aperiodic but recurring and not known as one-time phenomena. During these minima events, as the light decreases, the star becomes redder, then bluer, and the polarization increases. High dust column densities cause the initial dimming and reddening of light, while the bluing during minima and polarization is caused by an increase in the scattered light \citep{Grinin:1998, Waters:1998}. 

Some UXor stars such as SV Cep show long term periodicities in their light curves on the timescale of years \citep{Rostopchina:1999}. Even though UXor stars are usually early type, there are a few known late type stars that display UXor variations, such as the late K stars UY Aurigae \citep{Menard:1987}, AA Tau \citep{Bouvier:1999} and the M2 star DF Tau \citep{Chelli:1999}. Also, even though one of the explanations for the deep minima in UXor stars requires the disk to be edge on, the variations typically occur in pre-main sequence stars that are surrounded by circumstellar disks at an inclination of $45^{\circ}$ - $68^{\circ}$ \citep{Natta:2000}.

RW Aur A would be one of the rare late-type stars showing UXor variability. There are some apparent similarities in the long term light curve of the UXor star SV Cep when compared with our observations of RW Aur A. Both SV Cep and RW Aur A show minima lasting 100-400 days and long term changes in the median base line \citep{Rostopchina:1999}. Even in the extreme case of SV Cep, which experiences rare deep minima, it still shows 3 events in 36 years \citep{Rostopchina:1999}. We have examined the entire 50+ year light curve of RW Aur and find nothing resembling the depth and duration of the 2010-2011 dimming. Thus if RW Aur A is a UXor star, it would have the longest time lapse between minima known. 

An explanation of UXor minima is that they are the result of hydrodynamical fluctuations in either the inner rim or outer edge of the star's circumstellar disk \citep{Dullemond:2003}. Taking a conservative disk inclination for RW Aur A of $60^{\circ}$ (See figure \ref{fig_Disk_Geometry}), we determined in section \S5.2.3 that a feature at the edge of the RW Aur A disk, even with an extremely flared disk, would have to protrude an additional 15.5 AU from the disk to cross the star. A 15.5 AU protrusion is far larger then any known hydrodynamical fluctuations of a circumstellar disk. Therefore we explore the possibility that the disk of RW Aur A has a puffed-up inner rim causing the occultation. \citet{Dullemond:2003} determined that the height of the inner rim would be $H_{rim}$$\sim$0.2 $R_{rim}$ with a hydrodynamical fluctuation of $\sim$0.1 $R_{rim}$. We use the conservative inner radius of the RW Aur A disk of 0.103$\pm$0.005 AU from \citet{Eisner:2007}. This results in a maximum puffed-up inner rim height of 0.031 AU. Using the same inclination of the RW Aur disk, this would result in a relative height of 0.027 AU to our line of sight. At 0.103 AU from the star, in a disk inclined by $60^{\circ}$ , the height necessary to cross our line-of-sight is 0.0555 AU. Even using the most conservative values, the required height to cause extinction is still twice the maximum puffed-up height of the inner rim for the RW Aur A disk. Moreover, this puffing would need to have occurred only once for some unknown reason. However, the accretion rate onto the star during the occultation did not change appreciably (See \S4.2). Thus, it is unlikely that the dimming observed is the result of circumstellar extinction from a hydrodynamically puffed-up inner disk rim.

Unfortunately, we do not have any color or polarization observations of the system during the large dimming event and therefore cannot look for the reddening during the beginning of the dimming or color reversal and increased polarization of the light at the maximum depth. These types of observations would allow us to definitively determine if the 2010-2011 event is a UXor minimum. However, when comparing our observations to known UXor stars, if the 2010-2011 dimming was a UXor minima, it should have occurred more than once in the 50+ years of consistent observations. Also, RW Aur A does not fit the typical profile of a UXor star and therefore, we do not believe it to be a new UX Orionis star.

\section{\bf Summary and Conclusions}

The new observations of the RW Aur system from KELT-North show that a long ($\sim$180 days) deep ($\sim$2 magnitude) dimming occurred in late 2010 to early 2011. The event is also visible in the AAVSO archive which contains photometric observations of RW Aur. 

We have determined that the most plausible explanation for this event is that a fragment of the arm from the tidally disrupted circumstellar disk, thought to be caused by a recent fly-by of the B component, crossed the face of the A component. The observations from \citet{Cabrit:2006} show a tidally disrupted ``arm'' feature, 600 AU long, that is connected at the Northeast position of the A component and wraps around behind the star towards the B component. We calculate a maximum linear velocity of the occulter of 2.58 km s$^{-1}$, consistent with the maximum blue-shifted velocity, relative to RW Aur A, of the tidally disrupted arm, $\sim$3.1 km s$^{-1}$ \citep{Cabrit:2006}. Assuming Keplerian motion, the occulter is located at a distance of $\sim$180 AU from RW Aur A, which is over twice the maximum radius of the observed circumstellar disk, but could still be located in this tidal feature. 

The simulations performed by \citet{Clarke:1993} predict that the tidally disrupted arm, produced from a close stellar fly-by of a companion, would have a relative width of 0.05 the periastron distance. \citet{Cabrit:2006} calculated that the periastron distance, during the stellar fly-by of RW Aur B, would have been 100-140 AU. From our maximum linear velocity, we calculate the thickness of the arm fragment to be 0.27 AU ($\sim$0.003 the periastron distance). Evidently the tidal arm has remained fairly coherent despite the tidal disruption event. These observations may indicate that we are witnessing the leading edge of the tidally disrupted arm occulting the star, and that future occultations may arise from other portions of the tidal arm.  Therefore, we encourage observers to obtain multi-filter photometric observations of RW Aur in hopes of better characterizing future obscurations of the star.  These results also motivate additional detailed simulations to extend the early work of \citet{Clarke:1993} along with a reexamining of the spectra taken by \citet{Chou:2013} that coincide with the first half of the dimming event, including the point of maximum depth. Since the authors of \citet{Chou:2013} were unaware of the dimming event, a comparison of their results with out-of-occultation observations may show spectroscopic signatures of the occulting body. This rare observation provides insight into the dynamics of proto-planetary environments in binary star systems.

\vspace{0.2in}

We would like to thank David Weintraub, Nathan De Lee, Josh Winn, Catherine Espaillat, Cathie Clarke, John Johnson, and the entire KELT team for discussing with us our observations and analysis. 

We have used observational data from the SuperWASP photometric survey and the Wesleyan University Van Vleck \emph{UBVRI} photometric. We are thankful for the observations and data reduction performed.

Early work on KELT-North was supported by NASA Grant NNG04GO70G. J.A.P. and K.G.S. acknowledge support from the Vanderbilt Office of the Provost through the Vanderbilt Initiative in Data-intensive Astrophysics. This work has made use of NASAÕs Astrophysics Data System and the SIMBAD database operated at CDS, Strasbourg, France.

We acknowledge with thanks the variable star observations from the AAVSO International Database contributed by observers worldwide and used in this research. Specifically we would like to thank the AAVSO director Arne Henden. 

\maketitle

\end{document}